\documentclass[reprint,amsmath,amssymb,amsfonts,aps,floatfix]{revtex4-2}

\usepackage{natbib}
\usepackage{graphicx,color,rotating}
\usepackage[latin1]{inputenc}
\usepackage{textcomp}
\usepackage{dcolumn}
\usepackage{bm}     
\usepackage{upgreek}
\usepackage{xfrac}
\usepackage{array} 
\newcolumntype{L}[1]{>{\raggedright\let\newline\\\arraybackslash\hspace{0pt}}m{#1}}
\newcolumntype{C}[1]{>{\centering\let\newline\\\arraybackslash\hspace{0pt}}m{#1}}
\newcolumntype{R}[1]{>{\raggedleft\let\newline\\\arraybackslash\hspace{0pt}}m{#1}}

\usepackage{hyperref}					
\usepackage{xcolor}						
\hypersetup{			     			
    colorlinks,
    allcolors={blue}
}
\newcommand{\qmarkstt}[1]{{``\texttt{#1}''}}

\newcommand{\upspace}{\rule{0ex}{2.5ex}}

\newcommand{\notop}{{{}_{}}}

\newcommand{\mr}[1]{\ensuremath{\mathrm{#1}}}
\newcommand{\myvec}[1]{\bm{#1}}
\newcommand{\ee}{\mathrm{e}}
\newcommand{\ii}{\mathrm{i}}
\newcommand{\dm}{\mathrm{d}}
\newcommand{\avr}[1]{\big\langle #1 \big\rangle}
\newcommand{\avrsp}[1]{\avr{\!\avr{#1}\!}}

\DeclareMathOperator{\re}{Re}
\DeclareMathOperator{\im}{Im}

\newcommand{\iot}{{\ii\omega t}}
\newcommand{\miot}{{-\ii\omega t}}

\newcommand{\pp}{\partial}

\newcommand{\nablabf}{\boldsymbol{\nabla}}

\newcommand{\Lapl}{\nabla^2}

\newcommand{\grad}{\nablabf}

\renewcommand{\div}{\nablabf\cdot}

\newcommand{\scap}{\!\cdot\!}

\newcommand{\CCC}{\myvec{C}}

\newcommand{\eee}{\myvec{e}}

\newcommand{\Frad}{F^{\mathrm{rad}}}

\newcommand{\III}{\myvec{I}}

\newcommand{\nnn}{\myvec{n}}

\newcommand{\rrr}{\myvec{r}}

\newcommand{\sss}{\myvec{s}}

\newcommand{\uuu}{\myvec{u}}
\newcommand{\uun}{\myvec{u}}

\newcommand{\vvv}{\myvec{v}}
\newcommand{\vvn}{\myvec{v}}
\newcommand{\vnn}{{v}}

\newcommand{\zerovec}{\boldsymbol{0}}

\newcommand{\calE}{\mathcal{E}}

\newcommand{\cO}{c_0}

\newcommand{\cOfl}{c_0^\fl}

\newcommand{\Eac}{E_\mathrm{ac}}

\newcommand{\Eacfl}{E_\mr{ac}^\mr{fl}}
\newcommand{\Eacsl}{E_\mr{ac}^\mr{sl}}

\newcommand{\vol}{\mathcal{V}}

\newcommand{\deltan}{\delta}

\newcommand{\Omegafl}{{\Omega^\mr{fl}}}
\newcommand{\Omegasl}{{\Omega^\mr{sl}}}

\newcommand{\rhosl}{\rho^\mr{sl}}

\newcommand{\cOsqr}{c^{\,2_{}}_0}

\newcommand{\kapOfl}{\kappa_0^\fl}

\newcommand{\pO}{p_0}

\newcommand{\pI}{p_1}

\newcommand{\pII}{p_2}

\newcommand{\vvvO}{\vvv_0}

\newcommand{\vvvI}{\vvv_1}

\newcommand{\vvvII}{\vvv_2}

\newcommand{\rhoOfl}{\rho^\fl_0}

\newcommand{\rhoO}{\rho_0}

\newcommand{\rhoI}{\rho_1}

\newcommand{\SIcm}{\textrm{cm}}

\newcommand{\SIMHz}{\textrm{MHz}}

\newcommand{\SIkgm}{\textrm{kg}\:\textrm{m$^{-3}$}}
\newcommand{\SIkgpcm}{\SIkgm}
\newcommand{\SIm}{\textrm{m}}

\newcommand{\SImm}{\textrm{mm}}
\newcommand{\SImum}{\textrm{\textmu{}m}}
\newcommand{\SImu}{\textrm{\textmu{}}}
\newcommand{\SInm}{\textrm{nm}}

\newcommand{\SIkPa}{\textrm{kPa}}
\newcommand{\SIpTPa}{\textrm{TPa}^{-1}}

\newcommand{\SImPas}{\textrm{mPa}\:\textrm{s}}

\newcommand{\SIs}{\textrm{s}}

\newcommand{\SImps}{\SIm\,\SIs^{-1}}

\newcommand{\beq}[1]{\begin{equation} \eqlab{#1}}
\newcommand{\eeq}{\end{equation}}
\newcommand{\bsub}{\begin{subequations}}
\newcommand{\esub}{\end{subequations}}
\def\bal#1\eal{\begin{align}#1\end{align}}
\def\balat#1#2\ealat{\begin{alignat}{#1} #2 \end{alignat}}
\def\bsubal#1 #2\esubal{\bsuba{#1}\begin{align}#2\end{align} \esuba}     
\def\bsubalat#1#2#3\esubalat{\bsuba{#1} \begin{alignat}{#2} #3 \end{alignat} \esuba}
\newcommand{\bsuba}[1]{\bsub \eqlab{#1}}
\newcommand{\esuba}{\esub}

\newcommand{\eqlab}[1]{\label{eq:#1}}
\renewcommand{\eqref}[1]{Eq.~(\ref{eq:#1})}
\newcommand{\eqnoref}[1]{(\ref{eq:#1})}

\newcommand{\eqsref}[2]{Eqs.~(\ref{eq:#1}) and~(\ref{eq:#2})}

\newcommand{\figref}[1]{Fig.~\ref{fig:#1}}

\newcommand{\figsref}[2]{Figs.~\ref{fig:#1} and~\ref{fig:#2}}
\newcommand{\figlab}[1]{\label{fig:#1}}

\newcommand{\secref}[1]{Sec.~\ref{sec:#1}}

\newcommand{\secsref}[2]{Secs.~\ref{sec:#1} and~\ref{sec:#2}}
\newcommand{\seclab}[1]{\label{sec:#1}}
\newcommand{\tabref}[1]{Table~\ref{tab:#1}}

\newcommand{\tablab}[1]{\label{tab:#1}}

\newcommand{\sigmabf}{\bm{\sigma}}

\newcommand{\cL}{c_\mathrm{lo}}

\newcommand{\cT}{c_\mathrm{tr}}

\newcommand{\cLsqr}{c^2_\mathrm{lo}}
\newcommand{\cTsqr}{c^2_\mathrm{tr}}

\newcommand{\uuuI}{\myvec{u}_1}

\newcommand{\fl}{\mathrm{fl}}

\renewcommand{\sl}{\mathrm{sl}}

\definecolor{darkgreen}{rgb}{0.00, 0.50, 0.00}
\definecolor{DARKGREEN}{rgb}{0.00, 0.50, 0.00}
\definecolor{RED}{rgb}{1.00, 0.00, 0.00}
\definecolor{GREEN}{rgb}{0.00, 1.00, 0.00}
\definecolor{BLUE}{rgb}{0.00, 0.00, 1.00}
\definecolor{MAGENTA}{rgb}{1.00, 0.00, 1.00}

\newcommand{\kfl}{k^\mathrm{fl}}
\newcommand{\ksl}{k^\mr{sl}}
\newcommand{\SIGPa}{\textrm{GPa}}

\newcommand{\vIIray}{v_2^\mathrm{Rayl}}
\newcommand{\CCCeff}{\CCC^\mathrm{mm}}

\newcommand{\uuuM}{\uuu^\mathrm{mm}}
\newcommand{\rhoOeff}{\rho_0^\mathrm{mm}}

\begin{document}

\title{Enhanced quality factors at resonance in acoustofluidic cavities\\ embedded in matched elastic metamaterials}

\author{Valdemar Frederiksen}
\email{valde.freder@gmail.com} \affiliation{Department of Physics, Technical University of Denmark,\\
DTU Physics Building 309, DK-2800 Kongens Lyngby, Denmark}

\author{Henrik Bruus}
\email{bruus@fysik.dtu.dk}
\affiliation{Department of Physics, Technical University of Denmark,\\
DTU Physics Building 309, DK-2800 Kongens Lyngby, Denmark}

\date{29 October 2025}

\begin{abstract}
We show that by embedding liquid-filled acoustofluidic cavities in a metamaterial, the quality factor of the cavity at selected acoustic resonance modes can be enhanced by 2 to 3 orders of magnitude relative to a comparable conventional cavity by matching the coarse-grained elastic moduli of the metamaterial to the acoustic properties of the liquid.
\end{abstract}

\maketitle


\section{Introduction}
The quality factor $Q$ of an acoustic resonance mode in a liquid-filled acoustofluidic cavity embedded in an elastic solid is limited by the dissipation in the liquid, mainly due to the large stresses in the thin viscous boundary layers near the elastic walls of the cavity. For MHz ultrasound resonance modes in typical water-filled acoustofluidic cavities with kinematic viscosity $\nu$ at angular frequency $\omega$, $Q$-factors are often in the range 10 - 500 \cite{Barnkob2010}, a value set by the relevant length scale of the cavity relative to the thickness $\delta = \sqrt{\frac{2\nu}{\omega}}$ of the viscous boundary layer \cite{Bach2018}. For a long straight box-shaped cavity of length $L$, width $W$, and height $H$, with $H < W \ll L$, it is found that $Q = \frac{H}{\delta}$. Given a typical height $H = 160~\SImum$ and a standing half-wave along the width $W = 375~\SImum$ with a resonance frequency of 2~MHz, we have the boundary-layer width $\delta =  0.4~\SImum$, and thus $Q = 400$.

In the above example, the boundary layer is formed by the viscous friction in the liquid, as the acoustic velocity of the liquid is changing from its bulk value $v_1$ to zero at the nearly rigid wall over a distance of $\delta$. The smallness of $\delta$ is the reason that the dissipation $\dot{W}_\mr{BL} \approx \eta \big(\frac{v_1}{\delta}\big)^2$ in the boundary layer dominates the total dissipation in the system. The $Q$-factor may be increased considerably simply by removing the boundary layer. In this paper we show that the boundary layer may be removed or at least strongly suppressed by embedding the acoustic cavity not in a conventional elastic solid, denoted by superscript "$\sl$", but instead in an elastic metamaterial denoted by superscript "$\mr{mm}$". By tuning the coarse-grained elastic moduli of the metamaterial, it is possible at a given resonance to match at the solid-liquid interface of the cavity the vibrational velocity $\pp_t\uuu$ of the metamaterial to the acoustic velocity $\vvv_1$ of the liquid. This matching will suppress the boundary layer, and the dissipation of the system will then be limited by the small bulk dissipation of the liquid, and the $Q$-factor increases by a factor 200 to $\sim 10^5$.

\begin{figure}[b]
\centering
\includegraphics[width=0.9\columnwidth]{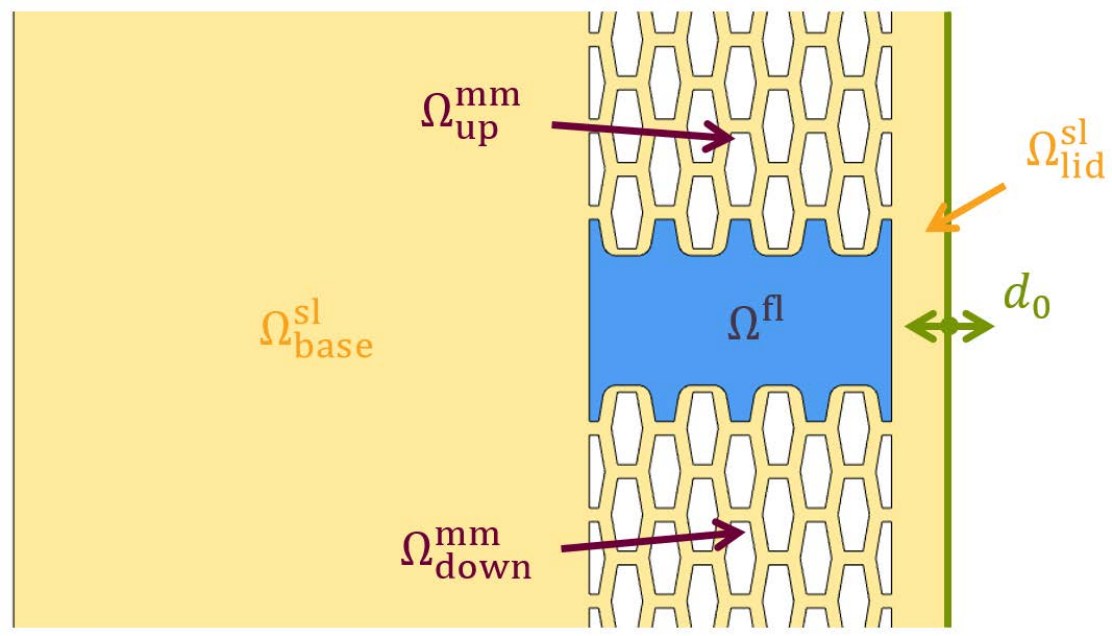}
\caption{\figlab{sketch} A sketch of the model system with its five domains: the solid base $\Omega^\sl_\mr{base}$, the fluid channel $\Omega^\fl$, the solid lid $\Omega^\sl_\mr{lid}$, and the metamaterial walls $\Omega^\mr{mm}_\mr{up}$ and $\Omega^\mr{mm}_\mr{down}$. The green double arrow represents the oscillation amplitude $d_0$ of the lid (green line).}
\end{figure}

The paper is organized as follows: In \secref{theory} we present the theory of acoustic cavities embedded in an elastic metamaterial. In \secref{model} we introduce a specific two-dimensional (2D) model, which is simulated numerically in \secref{opt}, using the finite-element software COMSOL Multiphysics, as well as analytically for a simplified case. Finally, in \secsref{discuss}{conclusion} we discuss the results and present our conclusions and outlook. Animations of selected results are provided in the Supplemental Material~\footnote{See Supplemental Material at \url{https://bruus-lab.dk/files/Frederiksen_metamaterial_acoustofluidics_suppl.zip} for animated gifs of \figsref{comparison}{focusing}.}.

\section{Theory}
\seclab{theory}
\subsection{The 2D model system}
Our proposed system is assumed to be translational invariant in the length direction $L$ and with the rectangular cross section sketched in \figref{sketch}. In this cross section a fluid-filled cavity $\Omega^\fl$ circumscribing a rectangle of width $W^\fl$ and height $H^\fl$ is surrounded by a solid material region $\Omega^\sl$ which is subdivided into four domains. The base $\Omega^\sl_\mr{base}$ is a large rectangular block of regular solid material of width $W^\sl$ and height $H^\sl$ and with a free surface to the left. The lid $\Omega^\sl_\mr{lid}$ is a thin rectangular region of regular solid material of width $W^\sl_\mr{lid}$ with a prescribed time-harmonic displacement $\uuu^\mr{phys} = d_0\cos(\omega t)\:\eee_y$ on its rightmost boundary. The remaining two rectangular regions $\Omega^\mr{mm}_\mr{up}$ and $\Omega^\mr{mm}_\mr{down}$ above and below the cavity consist of a mechanical metamaterial, where the small-scale geometry has been optimized such as to make the metamaterial move in sync with the fluid in the cavity. We treat all acoustic fields $g_1$ to be complex-valued with a harmonic time dependence $\ee^\miot$, such that the real part corresponds to the true physical field $g^\mr{phys}$,
\beq{g_phys}
 g^\mr{phys}(\rrr,t) = \mr{Re}[g_1(\rrr)\:\ee^\miot].
\eeq

\subsection{The fluid domain}

To model the acoustic behavior of the coupled fluid-solid system, we first consider the governing equations in the fluid domain $\Omega^\fl$ in terms of the pressure $p$, velocity $\vvv$, and density $\rho^\fl$, namely the continuity and Navier--Stokes equation,
 \bsubal{Navier--Stokes_full}
 \pp_t \rho^\fl &= - \div(\rho^\fl \vvv),
 \\
 \rho^\fl\pp_t\vvv  &= \div\sigmabf^\mathrm{fl} - (\rho^\fl\vvv\scap\nablabf)\vvv,
 \esubal
where the fluid stress tensor $\sigmabf^\mathrm{fl}$ is given in terms of $p$, $\vvv$, and the respective shear and bulk viscosity $\eta$ and $\eta^\mathrm{b}$,
 \beq{fluid stress}
 \sigmabf^\mathrm{fl} = \big[-p + \big(\eta^\mathrm{b}-\tfrac{2}{3}\eta\big) \div\vvv\big] \III
 +  \eta \big[\nablabf\vvv + (\nablabf\vvv)^\textsf{T}\big].
 \eeq
At the fluid-solid interface, which is the boundary $\pp\Omegafl$ of the fluid domain $\Omegafl$ with normal vector $\nnn$, the boundary conditions are continuity in velocity and normal stress,
 \bsubalat{BC_fl-sl}{2}
 \eqlab{no-slip}
 \vvv  &= \pp_t\uuu, &\;& \text{ at } \pp\Omegafl,
 \\
 \eqlab{stress_cont}
 \sigmabf^\mathrm{fl}\cdot \nnn &= \sigmabf^\sl \cdot \nnn, && \text{ at } \pp\Omegafl,
\esubalat
where $\sigmabf^\sl$ and $\pp_t\uuu$ is the stress and velocity of the material surrounding the fluid domain. We solve the above equations by applying perturbation theory in terms of the acoustic Mach number assuming fields to become progressively weaker with increasing orders. For the zeroth-order solution, we assume the fluid to be at rest $\vvvO = \zerovec$ with constant pressure $\pO$ and density $\rhoO^\fl$. We then assume the first-order fields to be time-harmonic, $g_1 = g_1(\rrr)\:\ee^\miot$, and relate the first-order pressure and density, $\pI(\rrr) = \cOsqr \rhoI^\fl$, by assuming a constant compressibility $\kapOfl = \big[\rhoOfl(\cOfl)^2\big]^{-1}$, where $\cOfl$ is the speed of sound. Finally, we truncate the perturbation expansion and keep only zeroth-, first-, and time-averaged second-order fields,
 \bsubal{rho_p_v_first}
 \rho^\fl(\rrr,t) &= \rhoO^\fl + \rhoO^\fl\kappa_0^\fl\pI(\rrr)\: \ee^{-\iot} + \rho_2^\fl(\rrr), \\
 p(\rrr,t) &= \pO + \pI(\rrr)\: \ee^{-\iot} + \pII(\rrr), \\
 \vvv(\rrr,t) &= \vvvI(\rrr)\: \ee^{-\iot} + \vvv_2(\rrr).
 \esubal
Inserting this expansion into \eqref{Navier--Stokes_full}, we obtain the first-order governing equations,
 \bsubal{Navier--Stokes_acoustic}
 i\omega\kappa_0^\fl\pI &= \div\vvvI,
 \\
 i\omega\rhoO^\fl\vvvI &= \nablabf \pI - \eta\Lapl \vvvI - \big(\eta^\mr{b}-\tfrac23\eta\big) \nablabf(\div\vvvI),
 \esubal
which we solve numerically, and which we work with analytically moving forward.

Numerically, we also solve for the second-order fields in $\Omegafl$, which are governed by the equations \cite{Bach2018},
 \bsubalat{Navier-Stokes-second}{2}
 \rhoO^\fl \div \vvvII + \div\avr{\rho_1^\fl \vvvI} &= 0,\;
 && \text{ in } \Omegafl,
 \\
 \div \sigmabf_2^\fl - \rhoO^\fl\div\avr{ \vvvI\vvvI} &= \zerovec,\;
 && \text{ in } \Omegafl,
 \\
 \vvvII + \avr{(\uuuI\cdot\nablabf)\vvvI} &= \zerovec,\;
 && \text{ at } \pp\Omegafl,
 \esubalat
where the bracket $\avr{A_1 B_1}$ represents the time average over one oscillation period of the product of any two complex-valued first-order fields $A_1$ and $B_1$,
  \beq{timeavrAB}
  \avr{A_1 B_1} = \frac12\re\big[A_1 B^*_1\big],
  \eeq
with the asterisk being the complex conjugate.

To characterize the utility of acoustic modes with regard to focusing microparticles by acoustophoresis, we calculate for each mode three useful quantities: The time-averaged acoustic energy density $\Eacfl$ in the fluid,
 \bsuba{CharacParam}
 \beq{Eac}
 \Eacfl = \frac{\kappa_0^\mathrm{fl}}{2}\avr{\pI\pI}
 \!+\! \frac{\rho_0^\fl}{2}\avr{\vnn_{1,i}\vnn_{1,i}}
 = \frac{\kappa_0^\mathrm{fl}}{4}|\pI|^2
 \!+\! \frac{\rho_0^\fl}{4}|\vvn_1|^2,
 \eeq
the Rayleigh streaming speed $\vIIray$ defined by,
 \beq{v_2_ray}
 \vIIray = \frac{3 \Eacfl}{2\rho_0^\fl \cOfl},
 \eeq
and the spatial average $\avrsp{v_2}$ of the magnitude of the streaming velocity $\vvvII$,
 \beq{v_2_mean}
 \avrsp{v_2}  = \frac{1}{\vol^\mathrm{fl}}\int_{\Omegafl} |\vvv_2| \:\dm V.
 \eeq
 \esuba

If we imagine placing a spherical particle with radius $a$ inside an acoustic cavity, the particle is subject to an acoustic radiation force $F_\mr{rad} \sim a^3 \kfl E_\mr{ac}^\fl$ pushing it towards the nearest pressure node (or anti-node) \cite{Gorkov1962, Bach2020, Winckelmann2023}. Simultaneously, a drag force $F_\mr{drag}\sim 6\pi a \eta v_2$ associated with the acoustic streaming field $\vvv_2$ will tend to instead pull this particle around in a vortex motion. Consequently, there exists a critical radius $a_\mr{c} \sim \:\delta\sqrt{v_2/\vIIray}$, for which $F_\mr{drag} \sim F_\mr{rad}$, and below which particle focusing ceases. For polystyrene particles in a the box-shaped 2-MHz cavity described in the introduction, this radius is roughly $a_\mr{c} \sim 1~\SImum$ \cite{Muller2012, Barnkob2012a}. For strong focusing of small particles one therefore needs $\Eacfl$ to be as high as possible while keeping the streaming field $\vvvII$ as low as possible.

\subsection{The solid domains}
In the solid domains, we apply linear elastodynamics to solve for the displacement field $\uuu(\rrr,t)$, such that the strain is $\sss = \frac12 (\nablabf \uuu + (\nablabf\uuu)^\textsf{T})$ and the Hookean stress $\sigmabf^\sl$ in terms of the stiffness tensor $\CCC$ is linear in  $\sss$,
 \beq{stress_tensor}
 \sigmabf^\sl = \CCC:\sss = C^\notop_{ijkl}s^\notop_{kl}.
 \eeq
The second equality employs index-notation, where summation over repeated indices is implied. For isotropic materials, $\CCC$ can be written in terms of the Voigt coefficients $C^\notop_{11}$ and $C^\notop_{44}$ as
 \beq{voigt}
 C^\notop_{ijkl} = (C^\notop_{11}-2C^\notop_{44})\deltan_{ij}\deltan_{kl} + C^\notop_{44}(\deltan_{ik}\deltan_{jl}+\deltan_{il}\deltan_{jk}).
 \eeq
Consequently, the governing equation for the time-harmonic displacement field $\uuu$ of the elastic solid is the Cauchy momentum equation,
 \bsubal{ElasSolid}
 \eqlab{uTimeHarmonic}
 \uuu(\rrr,t) & = \uuuI(\rrr)\: \ee^{-\iot},
 \\
 \eqlab{Cauchy_eq}
 -\omega^2\rho^\sl\: \uun_1 &= \div\sigmabf^\sl,\; \text{ in } \Omegasl.
 \esubal

\begin{figure}[t]
\centering
\includegraphics[width=0.9\columnwidth]{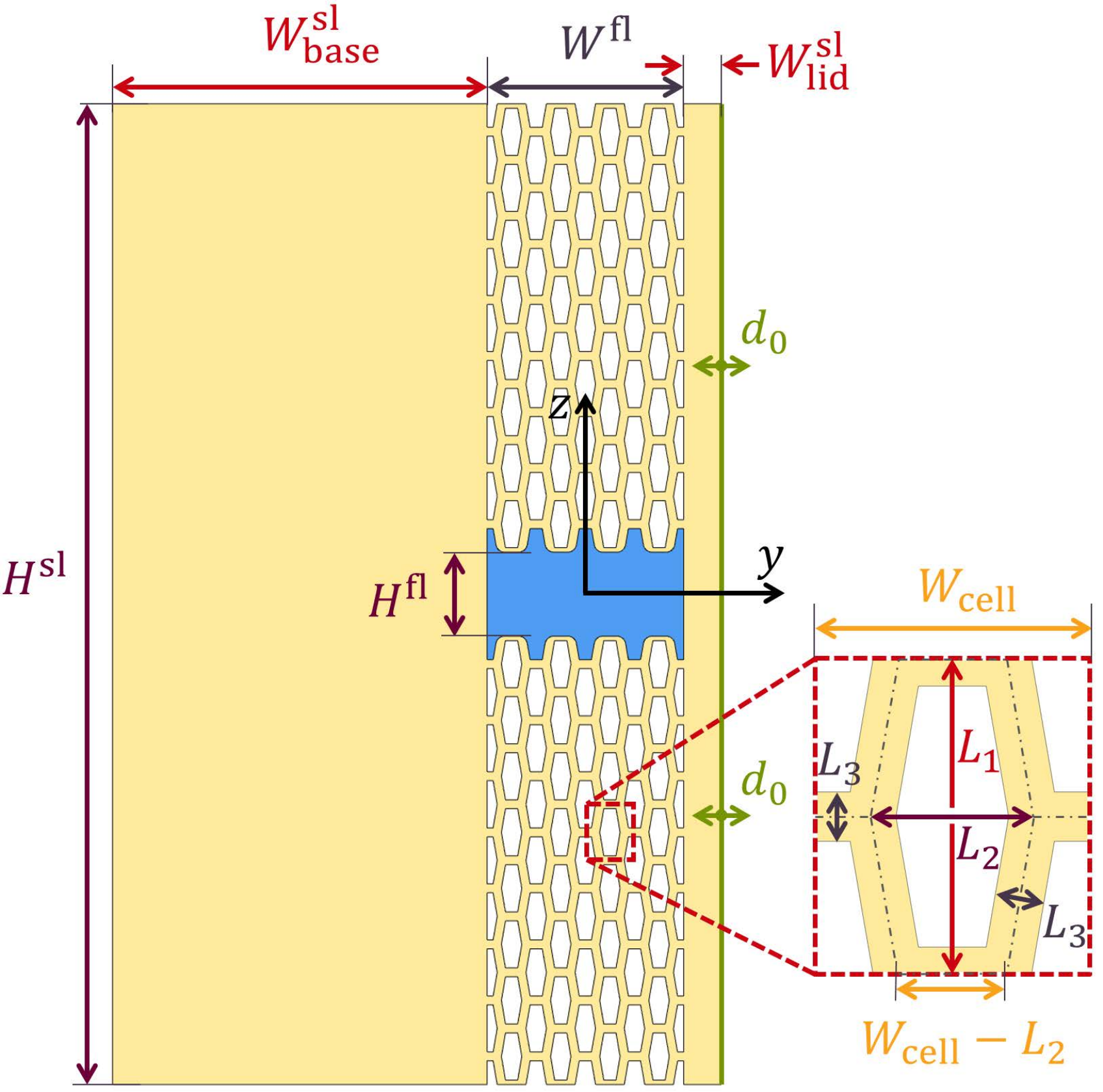}
\caption{\figlab{model} The geometry of the simulated device with two $(N_y\times N_z)=(4\times8)$ metamaterial arrays of elongated hexagonal unit cells on either side of the fluid channel. The device is actuated on the outer right-most surface (green line) with a time-harmonic uniform displacement amplitude of $d_0$ (green arrows), while all other walls are free to move having zero-stress conditions.}
\end{figure}

On the boundary of the solid domain, three types of boundary conditions apply: (1) continuity of velocity on the fluid-solid boundaries $\pp \Omega^\fl$, which supplement the continuity of stress \eqref{stress_cont}, (2) a prescribed time-harmonic displacement with amplitude $d_0$ on the actuated part $\pp\Omega_\mr{osc}^\sl$ of the outer boundary, and (3) zero stress on the remaining free part $\pp\Omega_\mr{free}^\sl$ of the outer boundaries,
 \bsubalat{BCsolid}{2}
 \eqlab{BCu1flsl}
 \uuuI & = \frac{\ii}{\omega}\;\vvn_1, && \text{ at } \pp\Omega^\fl,
 \\
 \eqlab{BCu1Osc}
 \uuuI & = d_0\; \nnn,\quad && \text{ at } \pp\Omega_\mr{osc}^\sl,
 \\
 \eqlab{BCzeroStress}
 \sigmabf^\sl \cdot \nnn & = \zerovec, &&\text{ at } \pp\Omega_\mr{free}^\sl.
 \esubalat

We end this section by defining the time-averaged acoustic energy density $\Eacsl$ within the solid as
 \beq{Esl}
 \Eacsl = \frac12 \avr{\big(C^\notop_{ijkl} s^\notop_{1,ij}\big) s_{1,kl}}
 + \frac12 \rhosl \omega^2 \avr{u_{1,i}u_{1,i}}.
 \eeq
The total time-averaged acoustic energy $\bar{U}_\mr{ac}$ of the device, the time-averaged acoustic power $\bar{P}_\mr{in}$ supplied to the device, and the quality factor $Q_n$ of the $n$th resonance modes at frequency $\omega_n$ of the device are thus,
 \bsubal{EtotPowerQ}
 \eqlab{Etot}
 \bar{U}_\mr{ac} &= \int_{\Omega^\fl} \Eacfl \: \dm V + \int_{\Omegasl} \Eacsl \: \dm V,
 \\
 \eqlab{Power}
 \bar{P}_\mr{in} & = \int_{\pp\Omegasl}  \avr{(-\ii\omega_n\uuu_1)
 \cdot(\sigmabf_1^\sl\cdot \nnn)} \:\dm A,
 \\
 \eqlab{Q}
 Q_n &= \frac{\omega_n\bar{U}_\mr{ac}}{\bar{P}_\mathrm{in}}.
 \esubal

\begin{table}[t]
\caption{\tablab{Params} Parameter values used in the model systems. For fused silica, the following parameters are computed from the values listed in Ref.~\cite{OndaCorpSolids}: $\alpha = 6.25\times 10^{-5}~\text{dB/cm} \times 0.115~\text{Np/dB} \times 100~\SIcm/\SIm$, $\gamma = \alpha \cL/\omega$,  $\re[C_{11}] = \rho\cLsqr$, $\re[C_{44}] = \rho\cTsqr$, and $\im[C_{ii}] = -2\gamma  \re[C_{ii}]$.}
\centering
\begin{ruledtabular}
    \begin{tabular}{lcrl}
    Parameter          & Symbol                 & Value                    & Unit      \\\hline
    \multicolumn{4}{l}{\upspace\textit{Fused silica} \cite{OndaCorpSolids}} \\
    Mass density       & $\rhosl$               & $2200$                   & $\SIkgpcm$\\
    Sound speed, longitudinal & $\cL$ & $5700$ & $\SIm/\SIs$ \\
    Sound speed, transverse & $\cT$ & $3750$ & $\SIm/\SIs$ \\
    Attenuation constant at 2~MHz & $\alpha$ & $714$ & $\SImu$Np/m \\
    Damping coefficient at 2~MHz & $\gamma$ & $0.324$ & ppm \\
    Elastic modulus 11, real part    & $\re[C^\notop_{11}]$   & $71.5$                   & $\SIGPa$  \\
    Elastic modulus 44, real part    & $\re[C^\notop_{44}]$   & $30.9$                   & $\SIGPa$  \\
    Elastic modulus 11, imag.~part  & $\im[C^\notop_{11}]$   & $-46.3$                  & $\SIkPa$  \\
    Elastic modulus 44, imag~part& $\im[C^\notop_{44}]$   & $-20.0$                  & $\SIkPa$  \\\hline
    \multicolumn{4}{l}{\upspace\textit{Water} \cite{Muller2014}} \\
    Mass density       & $\rho_0^\fl$       & $997.05$                 & $\SIkgpcm$\\
    Speed of sound     & $\cOfl$                & $1496.7$                 & $\SImps$  \\
    Dynamic viscosity  & $\eta$                 & $0.890$                  & $\SImPas$ \\
    Bulk viscosity     & $\eta^\mr{b}$          & $2.485$                  & $\SImPas$ \\
    Compressibility    & $\kappa_0^\mathrm{fl}$ & $447.7$                  & $\SIpTPa$ \\\hline
    \multicolumn{4}{l}{\upspace\textit{Geometry}, \figref{model}}\\
    Channel height     & $H^\fl$            & $160$                    & $\SImum$  \\
    Channel width      & $W^\fl$            & $375$                    & $\SImum$  \\
    Solid base width   & $W^\sl_\mr{base}$  & $714$                    & $\SImum$  \\
    Displacement amplitude  & $d_0$             & $0.1$                    & $\SInm$   \\
    \end{tabular}
\end{ruledtabular}
\end{table}

\begin{figure*}[t]
\centering
\includegraphics[width=0.9\textwidth]{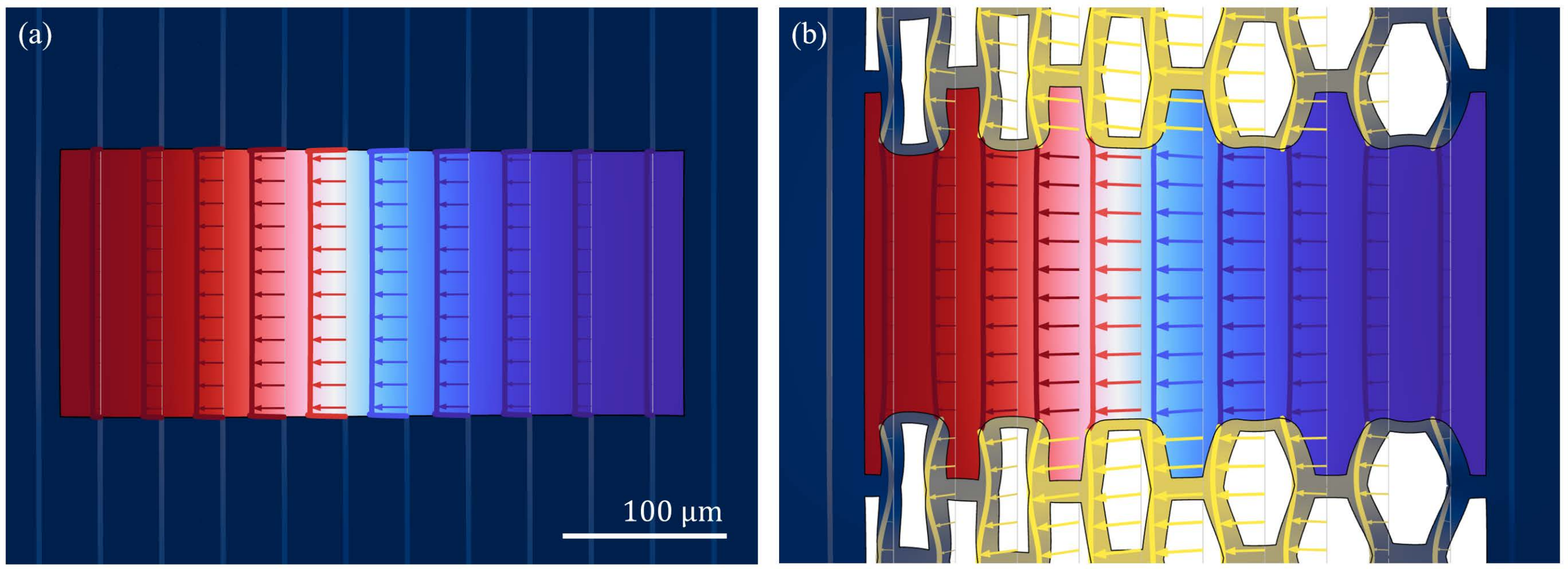}
\caption{\figlab{comparison} Simulation results in and near the fluid cavity: Color plots of the pressure field $\pI$ from min (blue) to max (red) in the fluid and of the magnitude $u_1$ from min (blue) to max (yellow) of the displacement field $\uuuI$ in the solid for (a) a conventional rectangular cavity and (b) the optimized cavity using the $5 \times 10$ metamaterial with the hexagonal unit cell defined in \figref{model}. The first-order displacements $\uuu_1$ and $\ii \omega^{-1}\vvvI$ of the solid and fluid are represented by evenly spaced, deformed lines and arrows. Animated versions of the panels are given in the Supplemental Material~\cite{Note1}.}
\end{figure*}

\subsection{The metamaterial domains}
The metamaterial domains consist of $N_y\! \times\! N_z$-arrays of the elongated hexagonal unit cell specified in \figref{model} and \tabref{Params}. Like the regular solid domains, the metamaterial domains are governed by the Cauchy equation. In a full numerical model this is implemented straightforwardly. However, when making analytical coarse-grained approximations, we describe a given metamaterial as a homogeneous anisotropic elastic solid governed by the Cauchy equation with effective values (superscript 'mm') of the mechanical fields, such as $\CCC \rightarrow \CCC^\mr{mm}$ and  $\uuu \rightarrow \uuu^\mr{mm}$. Most of the effective quantities are assumed to be local spatial average values of their exact counterparts with the single exception of $\CCC^\mr{mm}$, which instead is defined to satisfy the course-grained version of \eqref{stress_tensor} as follows. On a single unit cell (see the inset of \figref{model}) we a apply a constant strain $s_{yy} = d_0/W_\mr{cell}$ by prescribing a displacement of $\uuu = d_0 \eee_y$ on the rightmost edge $\calE_R$, a zero displacement $\uuu = \zerovec$ on the leftmost edge $\calE_L$, and the symmetry condition $u_z = 0$ on the top and bottom edges $\calE_T$ and $\calE_B$. Then the metamaterial values of the elastic coefficients are given by the respective average curve integrals as $C^\mr{mm}_{yyyy} = \frac{1}{H_\mr{cell}} \int_{\calE_L}\sigma_{yy}\dm \ell/s_{yy}$ and $C^\mr{mm}_{yyzz} = \frac{1}{W_\mr{cell}} \int_{\calE_T}\sigma_{zz}\dm \ell/s_{yy}$.

\section{Modelling the system}
\seclab{model}
\subsection{Materials and geometry of the model}

The model to be simulated numerically is defined as follows: The solid material is chosen to be fused silica glass due to its small attenuation constant, the fluid is water, and all parameter values are listed in \tabref{Params}.
The detailed cross-sectional geometry in the $y$-$z$ plane is shown in \figref{model}, and the model is assumed to be translational invariant in the out-of-plane $x$-direction. Note in particular the elongated hexagonal unit cells inspired by Ref.~\cite{Ling2020}, which initial configuration has been chosen to satisfy all the requirements for the course-grained elasticity and density derived later in this section. The design goal of the geometry is to support the first acoustic mode $n=1$, i.e.\ a standing half-wave resonance inside the rectangular fluid channel with a pressure node along the line $y=0$, and subsequently to fine-tune this mode to maximize its Q-factor $Q_1$. The obtained systems are then compared against a conventional reference system, where the metamaterial domains are replaced by solid glass.

\begin{table*}[t]
\caption{\tablab{Results} Simulation results for the reference system and nine different optimized metamaterial systems illustrated in \figref{comparison}.}
\centering
\begin{ruledtabular}
    \begin{tabular}{lcccccccccc}
    Parameter & Reference system & \multicolumn{9}{c}{Metamaterial systems}\\
    \hline
    $N_y$                  & - & 2  & 3  & 4  & 5  & 6  & 7  & 8  & 9  & 10 \\
    $N_z$                  & - & 4  & 6  & 8  & 10 & 12 & 14 & 16 & 18 & 20 \\
    $L_1~(\SImum)$         & - & 199.8& 138.6& 106.7& 86.3 & 72.1 & 62.0 & 54.4 & 48.4 & 43.6 \\
    $L_2~(\SImum)$         & - & 112.8 & 73.1 & 56.1 & 45.0 & 37.5 & 32.3 & 28.17 & 24.95 & 22.59 \\
    $L_3~(\SImum)$         & - & 34.0 & 22.63 & 16.98 & 13.58 & 11.33 & 9.70 & 8.49 & 7.54 & 6.79 \\
    $H^\sl~(\SImm)$    & 2.04 & 1.76 & 1.82 & 1.87 & 1.89 & 1.89 & 1.90 & 1.90 & 1.90 & 1.91 \\
    $f_\mr{res}~(\SIMHz)$  & 1.920 & 2.051 & 1.980 & 1.977 & 1.972 & 1.967 & 1.970 & 1.965 & 1.959 & 1.965 \\
    $Q/10^4$               & 0.043 & 4.3 & 6.6 & 8.7 & 10.4 & 12.2 & 13.8 & 15.2 & 16.2 & 17.2 \\
    $Q/Q^\mr{ref}$         & 1 & 99  & 153  & 200  & 240 & 283 & 321 & 351 & 375 & 397 \\
    $E_\mr{ac}^\fl~(\mr{MJm^{-3}})$ & $8.1\times 10^{-6}$ & 0.33 & 0.95 & 1.74 & 2.47 & 3.4 & 4.3 & 6.3 & 7.1 & 7.8 \\
    $\avrsp{v_2}/\vIIray$ & 0.282 & 0.053 & 0.047 & 0.041 & 0.032 & 0.030 & 0.0264 & 0.0234 & 0.0213 & 0.0198 \\
    \end{tabular}
\end{ruledtabular}
\end{table*}

\subsection{Implementation in COMSOL}
The governing equations and boundary conditions of \secref{theory} are, as in our previous work \cite{Bach2018, Bach2020, Winckelmann2023, Muller2012}, implemented in the \qmarkstt{Weak Form PDE Interface} of the commercial finite-element software COMSOL Multiphysics \cite{Comsol62}. Examples of numerical solutions are shown in \figref{comparison} for both a conventional rectangular fluid-filled channel embedded in an ordinary elastic solid and for a model with metamaterial.

\subsection{An approximate analytical solution}
As stated earlier, the goal of this paper is to create an acoustofluidic mode without viscous boundary layers so that $\vvvI$ is irrotational \cite{Bach2018}. In this case, the first-order governing equations reduce to a Helmholtz equation for $\pI$, while the velocity field $\vvvI$ is proportional to $\grad\pI$,
 \bsubal{HelmholtzGradp1}
 \eqlab{Helmholtz}
 \Lapl \pI &= -(1+\ii\gamma^\mathrm{fl})\frac{\omega^2}{(\cO^\fl)^2}\:\pI,
 \\
 \eqlab{Velocity}
 \vvvI &= \frac{-\ii}{\omega \rhoO^\fl}\nablabf\pI
 \esubal
where $\gamma^\mathrm{fl} = \omega\kappa_0^\fl(\eta^\mathrm{b}+\frac{4}{3}\eta)$ is the dimensionless bulk damping coefficient of the fluid, which for $\SIMHz$-frequencies in water is of order $10^{-5}$.
A mode is thus synchronized with the wall motion, if this curl-free bulk velocity along the entire boundary of the fluid domain satisfies the no-slip condition~\eqnoref{no-slip}, which in terms of $\pI$ is
 \beq{synchronization}
 \frac{1}{\omega^2 \rhoO^\fl}\nablabf\pI = \uuuI,\; \text{ at } \pp\Omegafl.
 \eeq
This condition gives us a strong hint about how to construct a cavity with a wall-synchronized resonance mode. We start by assuming that the cavity contains an ideal standing pressure half-wave along the $y$-direction,
 \beq{pressure}
 \pI = \pI^0 \sin(\kfl y),\; \text{ with }\; \kfl = \frac{\pi}{W^\fl}.
 \eeq
Neglecting $\gamma^\mathrm{fl}$, this mode obeys \eqref{Helmholtz} for $\omega = \cO^\fl k^\fl$. Next, to satisfy the wall-synchronization condition~\eqnoref{synchronization}, the displacement $\uuuM_1$ in the two metamaterial regions $\Omega^\mr{mm}_\mr{up}$ and $\Omega^\mr{mm}_\mr{down}$ may be chosen to be,
 \beq{meta_displacement_inside}
 \uuuM_1 = \frac{\pI^0 \kfl}{\omega^2 \rhoO^\fl}\;\cos(\kfl y)\:\eee_y ,\; \text{ in }\; \Omega^\mr{mm}_\mr{up} \;\text{ and }\; \Omega^\mr{mm}_\mr{down}.
 \eeq
Finally, one way to ensure that the displacement field $\uun_1$ in the isotropic base solid $\Omegasl$
has a vanishing amplitude along the base-metamaterial interface at $y = -\frac12 W^\fl$, while maintaining a non-vanishing normal stress, is to assume that $\uun_1$ takes the form of a standing longitudinal displacement wave with a node at $y = -\frac12 W^\fl$,
 \beq{meta_displacement_outside}
 \uun_1 = u_1^0\; \sin\big[\ksl (y+\tfrac12 W^\fl)\big]\;\eee_y,\;
 \text{ in }\; \Omega^\sl_\mr{base}.
 \eeq
From the Cauchy equation~\eqnoref{Cauchy_eq} and the stress continuity boundary conditions \eqsref{stress_cont}{BCzeroStress}, this assumption results in the amplitude and wavenumber relations,
 \bsubal{base_amplitude}
 u_1^0 &= \frac{\pI^0}{k^\sl C^\notop_{11}},\\
 k^\sl &= k^\fl\cOfl\sqrt{\frac{\rho^\sl}{C^\notop_{11}}}
 = k^\fl\frac{\cOfl}{\cO^\mr{sl,L}},
 \esubal
the quarter-wave width condition,
 \beq{quarter_wave}
 W^\sl_\mr{base} = \frac{\pi}{2\ksl} = \frac{\cO^\mr{sl,L}}{2 \cOfl} W^\fl,
 \eeq
and the following set of conditions on the coarse-grained metamaterial parameters $\CCCeff$ and $\rhoOeff$,
 \beq{mmconditions}
 C^\mr{mm}_{iyyy} = \frac{1}{\kappa_0^\fl}\deltan_{iy}, \quad
 C^\mr{mm}_{izyy} = \frac{1}{\kappa_0^\fl}\deltan_{iz}, \quad
 \rhoOeff = \rhoO^\fl.
 \eeq
In the limit where the lid thickness $W^\sl_\mr{lid}$ tends to zero, and the device height $H^\sl$ tends to infinity, this system has a near-perfectly synchronized resonance mode at frequency $f = \cO^\fl/W^\fl$. In the following section we show that even in the case of a geometry with finite values of $W^\sl_\mr{lid}$ and $H^\sl$, the above conditions serve as en excellent starting point for numerical determination of metamaterials that support well-synchronized modes.

\section{Numerical simulation}
\seclab{opt}
We now present the main result of our work: the optimization of the metamaterial unit cell to maximize the $Q$ value of the acoustic cavity.

\subsection{Optimizing the system for maximum Q value}
\seclab{numerics}
Our numerical optimization procedure for the metamaterial is divided into two steps. First, we find a metamaterial which satisfies the three criteria listed in \eqref{mmconditions}. For this, we opted for a hexagonal metamaterial, which has shown high tuneability in previous work \cite{Ling2020}, and which due to its twofold mirror-symmetry already satisfies the criteria $C_{iyyy}^\mr{mm}=C_{yyyy}^\mr{mm}\deltan_{iy}$ and $C_{izyy}^\mr{mm}=C_{zzyy}^\mr{mm}\deltan_{iz}$. By simulation in COMSOL Multiphysics of one unit cell in the metamaterial, we calculate the course-grained stiffness tensor $\CCC^\mr{mm}(L_1,L_2,L_3)$ and manually adjust the parameters $L_1$, $L_2$, and $L_3$ defined in \figref{model} until \eqref{mmconditions} is satisfied exactly. In our case with fused silica, this step results in the parameters
 \beq{Li_first}
 L_1 = \frac{412\, \SImum}{N_y} \quad L_2 = \frac{206.6\, \SImum}{N_y} \quad L_3 = \frac{67.9\, \SImum}{N_y}.
 \eeq

The second step in the optimization procedure involves simulating the full system repeatedly using the built-in COMSOL \qmarkstt{Optimization Interface}. Starting with the parameters found in the first step, and then for each simulation changing the height $L_1$ and the hole-width $L_2$, the $Q$-factor of the resonance mode is maximized. The resulting set of optimized values of $L_1$ and $L_2$ are listed in \tabref{Results}.

\subsection{Energy density and Q-factor}
In \figref{comparison}, the conventional reference cavity is compared to the optimized metamaterial cavity with $(N_y\times N_z)=(5\times 10)$ unit cells. For the conventional cavity, the velocity amplitude of the acoustic mode in the bulk of the cavity away from the boundary layers is far greater than the wall velocity, leading to large velocity gradients inside the boundary layers near the fluid-solid interface. These gradients are far smaller inside the optimized metamaterial cavity, because there the first-order displacements $\uuu_1$ and $i\omega^{-1}\vvv_1$ are synchronized across the interface. This difference has a considerable effect on the $Q$-factor of the mode, which in the following is always the $n=1$ mode, so henceforth we drop the mode index: The conventional cavity has $Q=432$, whereas the optimized ($5\times 10$)-metamaterial cavity has $Q=1.04\times 10^5$, which is nearly 250 times greater.

\begin{figure}[b]
\centering
\includegraphics[width=0.92\columnwidth]{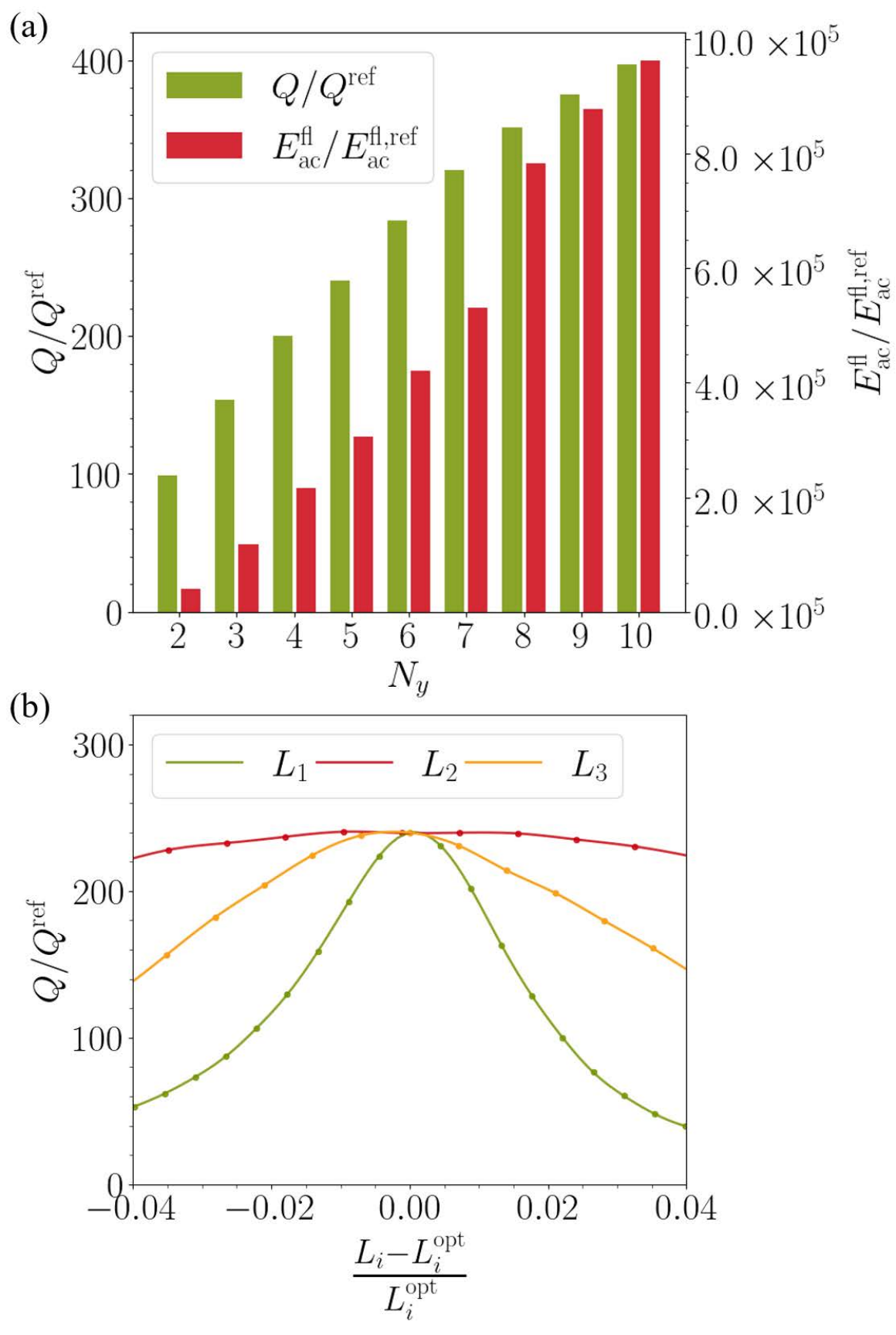}
\caption{\figlab{graphs} (a) Simulated $Q$-factors and acoustic energy density $E_\mr{ac}^\fl$ relative to the all-solid reference system (superscript "ref") for increasing numbers of unit cells $N_y$ across the metamaterial regions. (b) The  normalized $Q$ factor $Q/Q^\mr{ref}$ of the $N_y=5$ cavity plotted against the relative deviation of either $L_1$, $L_2$, or $L_3$ from the optimum.}
\end{figure}
\begin{figure*}[t]
\centering
\includegraphics[width=0.9\textwidth]{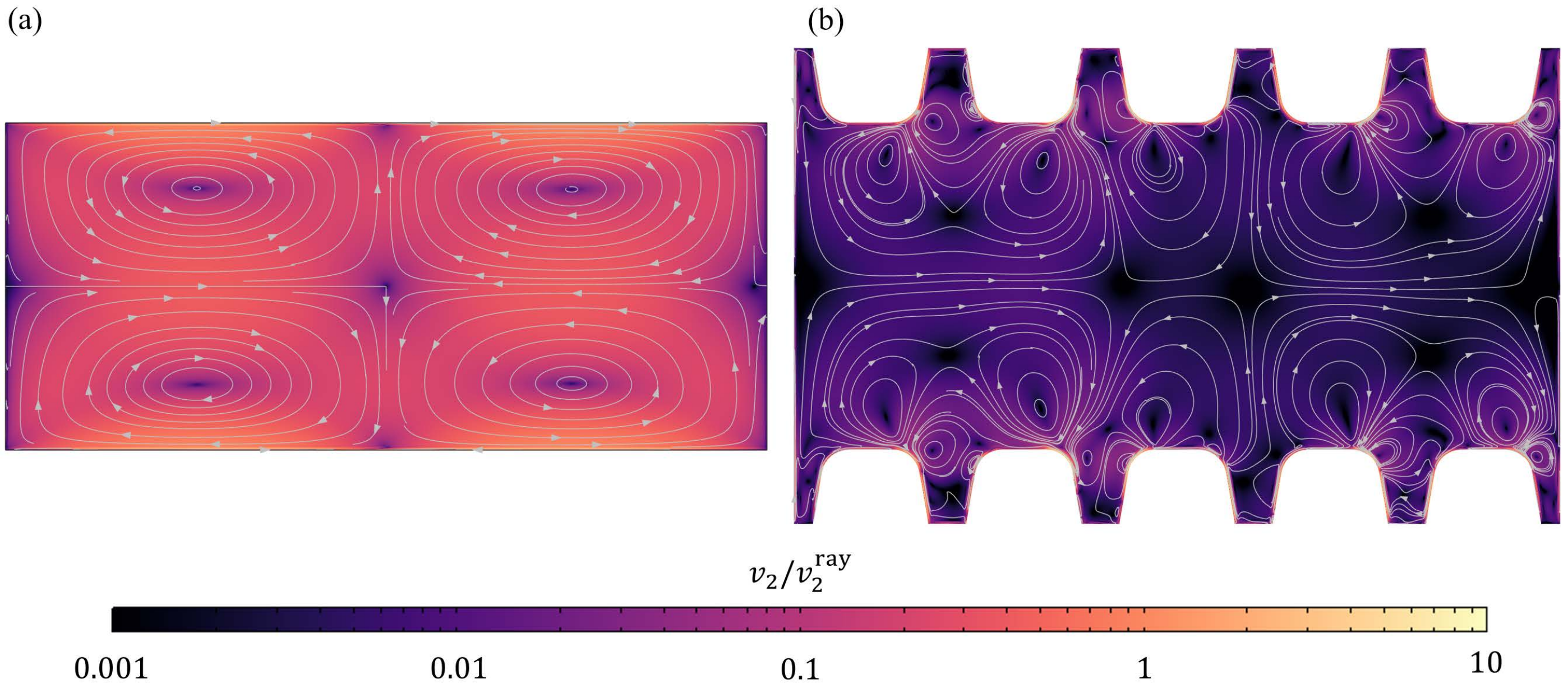}
\caption{\figlab{streaming} Streaming fields $\vvv_2$ normalized by the Rayleigh streaming speed $\vIIray$ inside (a) the reference cavity and (b) the $(N_y, N_z) = (5,10)$ synchronized metamaterial cavity.}
\end{figure*}

In \figref{graphs}(a) the acoustic energy density $\Eacfl$ and the $Q$-factor of the optimized systems are plotted against the number of unit cells $N_y$ across the channel. For $N_y=2$ the optimized systems already outperform the reference system for both metrics, and these performance metrics increase monotonically with $N_y$. For $Q$, this relationship is initially linear, whereas $\Eac^\fl$ grows quadratically. At higher values of $N_y$, these increases taper off.

To illustrate the sensitivity of the optimized resonance modes to the geometry parameters, we show in \figref{graphs}(b) how the $Q$-factor of the ($5\times 10$)-metamaterial cavity mode drops when $L_1$, $L_2$, and $L_3$ deviate from their optimized values. The greatest relative sensitivity is associated with $L_1$, where a $1.7\%$ deviation from optimum ($1.5~\SImum$ away from $86.3~\SImum$) halves the $Q$-factor of the cavity. The greatest absolute sensitivity is however due to $L_3$, where a change of only $0.5~\SImum$ away from the $13.58~\SImum$ optimum halves the $Q$-factor of the cavity.

\begin{figure*}[t]
\centering
\includegraphics[width=0.9\textwidth]{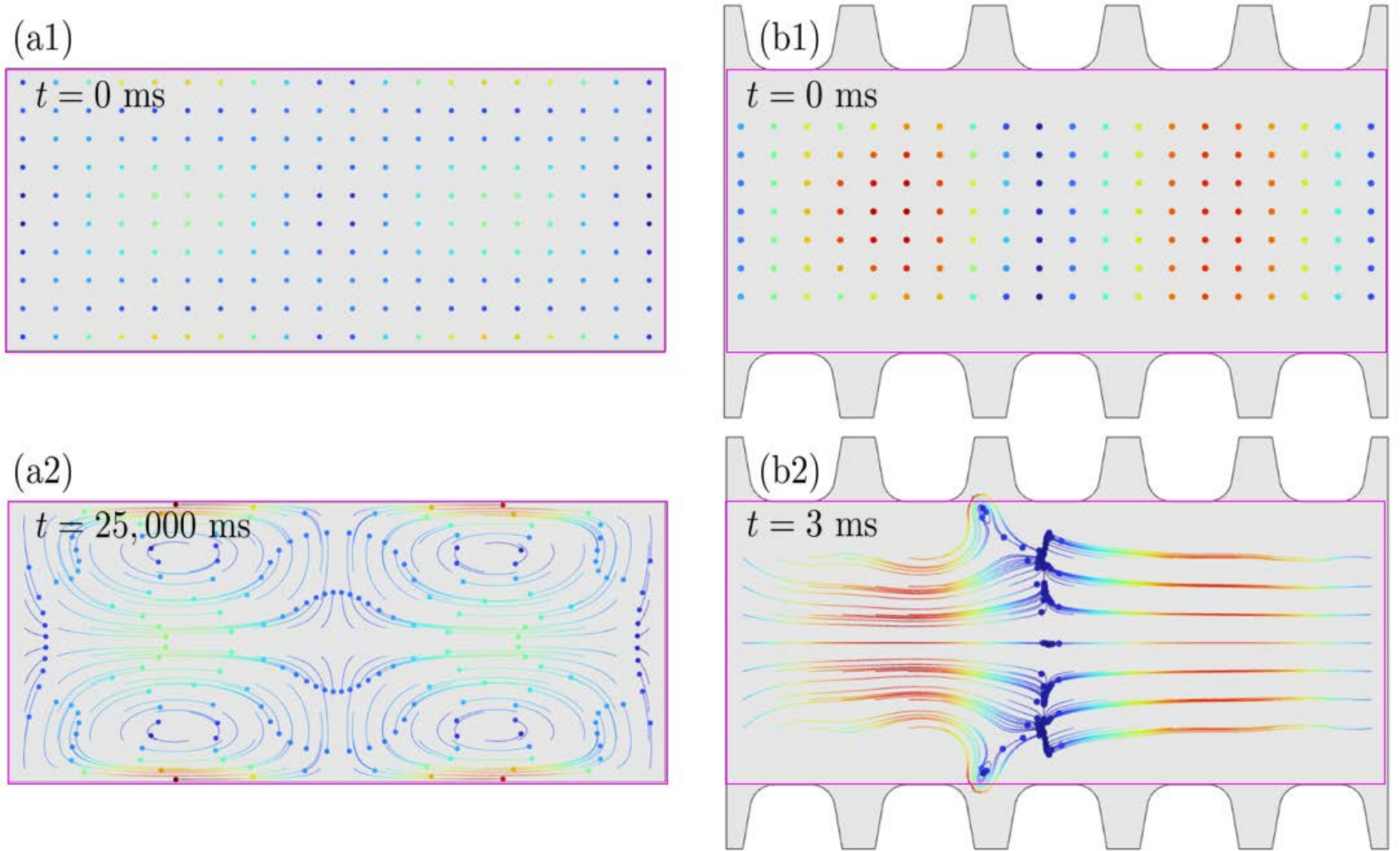}
\caption{\figlab{focusing} COMSOL simulation of the acoustophoretic trajectories of 250-nm-radius polystyrene  nanoparticles (dots) with initial positions in a regular grid. (a1) The reference device at $t = 0$~ms with the initial velocities set equal to the local streaming velocity represented by colors from 0 (dark blue) to $7~\SImum/\SIs$ (dark red). (a2) The same as panel (a1) but at $t = 25,000$~ms and showing the individual trajectories (lines) colored according to speed along the trajectories using the same color scheme as in panel (a1). (b1) same as panel (a1) at $t = 0$~ms, but for the metamaterial devices and with the velocity color from 0 (dark blue) to $130,000~\SImum/\SIs$ (dark red). (b2) The same as panel (b1) but for $t = 3$~ms and showing the individual trajectories (lines) colored according to speed along the trajectories using the same color scheme as in panel (b1). Animated versions of the panels are given in the Supplemental Material~\cite{Note1}.}
\end{figure*}

\subsection{Streaming}
\seclab{streaming}
To better understand the usefulness of the synchronized metamaterial cavities for doing acoustophoresis, we study the acoustic streaming field $\vvv_2$, as this will put a lower bound on the size of the particles that can be focused inside the cavity. For this reason, the streaming field of the optimized $(N_y \times N_z) = (5\times 10)$ metamaterial cavity is compared to that of the reference cavity. On one hand, one would expect the synchronized cavity to produce less streaming than the reference cavity, since its boundary layers are considerably weaker. On the other hand, the many corners along the fluid-solid interface of the synchronized cavity tend to amplify the streaming \cite{Ovchinnikov2014}. That the former effect dominates is revealed by  \tabref{Results}, where it is seen that the normalized streaming $\avrsp{v_2}/\vIIray$ is monotonically suppressed (nearly $\propto N_y^{-1}$) by more than an order of magnitude when going from the reference cavity up to the $(N_y \times N_z) = (10\times20)$ synchronized metamaterial cavity. Such a $10$-fold reduction in the streaming speed implies a $\sqrt{10}$-fold decrease in the critical radius $a_c$ from $\sim 1~\SImum$ to $\sim 0.3~\SImum$, allowing significantly smaller particles to be focused.

We note in \figref{streaming} that the streaming field of the synchronized metamaterial cavity is not only quantitatively, but also qualitatively different from that in the reference cavity. Whereas the streaming field in the reference cavity is dominated by the conventional four large rolls filling up the entire cross section, the streaming field in the synchronized cavity instead contains many smaller streaming vortices of various shapes and sizes, resulting in a less uniform streaming pattern, but clearly with more suppression of the streaming $\avrsp{v_2} \sim \frac{1}{50} \vIIray$ in the middle third of the channel, $-\frac16 H^\fl < z < \frac16 H^\fl$, such that $a_c$ drops down to $\sim 0.15~\SImum$ in this region.

\subsection{Particle focusing}

A remarkable addition to the discussion of the critical particle size $a_c$ is that \tabref{Results} reveals how the fluid pressure field $p_1^\mr{mm}$ in the synchronized metamaterial device scales with $p_1^\mr{ref}$ of the reference device and the Q-factors, $p_1^\mr{mm} \sim (Q^\mr{mm}/Q^\mr{ref})\:p_1^\mr{ref}$. This relation implies that the acoustic radiation force $\Frad$ on a suspended particle is a factor
$(Q^\mr{mm}/Q^\mr{ref})^2 \sim 4\times 10^4$ larger in the synchronized device than in the reference device for the same actuation amplitude. Consequently, the focusing time $\tau \propto \frac{\eta (W^\fl)^2}{a^2 \Eacfl}$ \cite{Barnkob2012}, or
$\tau^\mr{mm} \propto \left(\frac{Q^\mr{ref}}{Q^\mr{mm}}\right)^2\:\frac{\eta (W^\fl)^2}{a^2 E_\mr{ac}^{\fl,ref}}$, is $4\times10^4$ times faster in the former device than in the latter for a given particle size. Alternatively, a particle two orders of magnitude smaller, $a^\mr{mm} \sim 0.01 a^\mr{ref}$ will focus at the same time in the metamaterial device as $a^\mr{ref}$ will do in the reference device at the same actuation amplitude.

The dramatically improved nanoparticle acoustophoresis in the metamaterial device compared to the reference device is illustrated in \figref{focusing}. Here, the acoustophoretic velocity of 250-nm-radius nanoparticles changes by a factor of nearly $2\times10^4$ from $7~\SImum/\SIs$ to $130,000~\SImum/\SIs$. Moreover, where no focusing is observed in the reference device even on the long time scale of 25,000~ms, the 250-nm-radius nanoparticles are focused in the metamaterial device near the vertical pressure node in the channel center in just 3~ms.

\section{Discussion}
\seclab{discuss}

The fraction of acoustic energy stored inside the fluid $\bar{U}^\fl_\mr{ac}/\bar{U}_\mr{ac}\approx 0.2$ for the synchronized cavities. With the bulk damping factors $\gamma^\fl\approx 2\times 10^{-5}$ and $\gamma^\sl\approx 6\times 10^{-7}$, the theoretical upper bound for $Q$ should therefore be\cite{Hahn2015}
 \beq{bulk_damping}
 Q^\mr{bulk} = \frac{\bar{U}_\mr{ac}}{\gamma^\fl \bar{U}^\fl_\mr{ac}+ \gamma^\sl\bar{U}^\sl_\mr{ac}} \approx 2.2\times 10^5.
 \eeq

In \tabref{Results} we see that the highly resolved cavities tend to approach this theoretical upper bound quite nicely, and we may thus conclude that this synchronization manages to remove the dissipation associated with the viscous boundary layers rather reliably.

Looking ahead, if we were to speculate how one may reduce this dissipation even further and thus obtain still higher $Q$-factors, theoretically it should be possible to construct acoustic modes in such a way that the fluid contains less than a half-wave across the channel at resonance, but still contains a pressure node. This would greatly lower the bulk dissipation within the fluid.

We note that the proposed cavities have a harder time suppressing the streaming velocities compared to how easily $Q$-factors are raised. As mentioned earlier, we attribute this to be caused by the many corners around the fluid channel \cite{Ovchinnikov2014}, even though these corners have been rounded a bit to avoid generating too much streaming. One may also attempt to construct a cavity with fewer convex corners, or apply one of the methods proposed for suppression of acoustic streaming in previous work \cite{Bach2020, Karlsen2018, Winckelmann2021}. Clearly, there are many ways to suppress the remaining streaming further.

With regard to how such synchronized cavities might be fabricated, one possibility is to  rely on drawing towers such as the ones used for the fabrication of photonic crystal optical fibers. We note that such drawing towers already are designed to shape fused silica glass with a feature length scale lower than the length scales considered in this work \cite{Tajima2004}. Another possibility is to employ two-photon 3D-printing and fabricate the metamaterial structure directly, a possibility that has been demonstrated recently on comparable intricate micro- and nanoscale structures in glass \cite{Bauer2023}.

\section{Conclusion}
\seclab{conclusion}

By using metamaterials in acoustofluidic systems, it is possible to reduce the acoustic boundary layers dramatically by synchronizing the motion of the fluid to that of the adjacent material. This will in turn remove the viscous friction near the fluid-solid interface, and it may also, but to a lesser extent, reduce the acoustic streaming. This technique of using acoustic metamaterials to remove viscous friction could help increase $Q$-factors in a wider range of microelectromechanical systems. Specifically for acoustophoretic focusing of particles, synchronized metamaterial cavities may lead to faster focusing of smaller submicron particles compared to the current state-of-the-art of using cavities in conventional materials.

%
%


\begin{thebibliography}{19}%
\makeatletter
\providecommand \@ifxundefined [1]{%
 \@ifx{#1\undefined}
}%
\providecommand \@ifnum [1]{%
 \ifnum #1\expandafter \@firstoftwo
 \else \expandafter \@secondoftwo
 \fi
}%
\providecommand \@ifx [1]{%
 \ifx #1\expandafter \@firstoftwo
 \else \expandafter \@secondoftwo
 \fi
}%
\providecommand \natexlab [1]{#1}%
\providecommand \enquote  [1]{``#1''}%
\providecommand \bibnamefont  [1]{#1}%
\providecommand \bibfnamefont [1]{#1}%
\providecommand \citenamefont [1]{#1}%
\providecommand \href@noop [0]{\@secondoftwo}%
\providecommand \href [0]{\begingroup \@sanitize@url \@href}%
\providecommand \@href[1]{\@@startlink{#1}\@@href}%
\providecommand \@@href[1]{\endgroup#1\@@endlink}%
\providecommand \@sanitize@url [0]{\catcode `\\12\catcode `\$12\catcode
  `\&12\catcode `\#12\catcode `\^12\catcode `\_12\catcode `\%12\relax}%
\providecommand \@@startlink[1]{}%
\providecommand \@@endlink[0]{}%
\providecommand \url  [0]{\begingroup\@sanitize@url \@url }%
\providecommand \@url [1]{\endgroup\@href {#1}{\urlprefix }}%
\providecommand \urlprefix  [0]{URL }%
\providecommand \Eprint [0]{\href }%
\providecommand \doibase [0]{https://doi.org/}%
\providecommand \selectlanguage [0]{\@gobble}%
\providecommand \bibinfo  [0]{\@secondoftwo}%
\providecommand \bibfield  [0]{\@secondoftwo}%
\providecommand \translation [1]{[#1]}%
\providecommand \BibitemOpen [0]{}%
\providecommand \bibitemStop [0]{}%
\providecommand \bibitemNoStop [0]{.\EOS\space}%
\providecommand \EOS [0]{\spacefactor3000\relax}%
\providecommand \BibitemShut  [1]{\csname bibitem#1\endcsname}%
\let\auto@bib@innerbib\@empty
\bibitem [{\citenamefont {Barnkob}\ \emph {et~al.}(2010)\citenamefont
  {Barnkob}, \citenamefont {Augustsson}, \citenamefont {Laurell},\ and\
  \citenamefont {Bruus}}]{Barnkob2010}%
  \BibitemOpen
  \bibfield  {author} {\bibinfo {author} {\bibfnamefont {R.}~\bibnamefont
  {Barnkob}}, \bibinfo {author} {\bibfnamefont {P.}~\bibnamefont {Augustsson}},
  \bibinfo {author} {\bibfnamefont {T.}~\bibnamefont {Laurell}},\ and\ \bibinfo
  {author} {\bibfnamefont {H.}~\bibnamefont {Bruus}},\ }\bibfield  {title}
  {\bibinfo {title} {Measuring the local pressure amplitude in microchannel
  acoustophoresis},\ }\href {https://doi.org/10.1039/b920376a} {\bibfield
  {journal} {\bibinfo  {journal} {Lab Chip}\ }\textbf {\bibinfo {volume}
  {10}},\ \bibinfo {pages} {563} (\bibinfo {year} {2010})}\BibitemShut
  {NoStop}%
\bibitem [{\citenamefont {Bach}\ and\ \citenamefont {Bruus}(2018)}]{Bach2018}%
  \BibitemOpen
  \bibfield  {author} {\bibinfo {author} {\bibfnamefont {J.~S.}\ \bibnamefont
  {Bach}}\ and\ \bibinfo {author} {\bibfnamefont {H.}~\bibnamefont {Bruus}},\
  }\bibfield  {title} {\bibinfo {title} {Theory of pressure acoustics with
  viscous boundary layers and streaming in curved elastic cavities},\ }\href
  {https://doi.org/10.1121/1.5049579} {\bibfield  {journal} {\bibinfo
  {journal} {J. Acoust. Soc. Am.}\ }\textbf {\bibinfo {volume} {144}},\
  \bibinfo {pages} {766} (\bibinfo {year} {2018})}\BibitemShut {NoStop}%
\bibitem [{Note1()}]{Note1}%
  \BibitemOpen
  \bibinfo {note} {See Supplemental Material at \protect \url
  {https://bruus-lab.dk/files/Frederiksen_metamaterial_acoustofluidics_suppl.zip}
  for animated gifs of Figs.~\ref {fig:comparison} and~\ref
  {fig:focusing}.}\BibitemShut {Stop}%
\bibitem [{\citenamefont {Gorkov}(1962)}]{Gorkov1962}%
  \BibitemOpen
  \bibfield  {author} {\bibinfo {author} {\bibfnamefont {L.~P.}\ \bibnamefont
  {Gorkov}},\ }\bibfield  {title} {\bibinfo {title} {On the forces acting on a
  small particle in an acoustical field in an ideal fluid},\ }\href@noop {}
  {\bibfield  {journal} {\bibinfo  {journal} {Sov. Phys.--Dokl.}\ }\textbf
  {\bibinfo {volume} {6}},\ \bibinfo {pages} {773} (\bibinfo {year} {1962})},\
  \bibinfo {note} {[Doklady Akademii Nauk SSSR \textbf{140}, 88
  (1961)]}\BibitemShut {NoStop}%
\bibitem [{\citenamefont {Bach}\ and\ \citenamefont {Bruus}(2020)}]{Bach2020}%
  \BibitemOpen
  \bibfield  {author} {\bibinfo {author} {\bibfnamefont {J.~S.}\ \bibnamefont
  {Bach}}\ and\ \bibinfo {author} {\bibfnamefont {H.}~\bibnamefont {Bruus}},\
  }\bibfield  {title} {\bibinfo {title} {Suppression of acoustic streaming in
  shape-optimized channels},\ }\href
  {https://doi.org/10.1103/PhysRevLett.124.214501} {\bibfield  {journal}
  {\bibinfo  {journal} {Phys. Rev. Lett.}\ }\textbf {\bibinfo {volume} {124}},\
  \bibinfo {pages} {214501} (\bibinfo {year} {2020})}\BibitemShut {NoStop}%
\bibitem [{\citenamefont {Winckelmann}\ and\ \citenamefont
  {Bruus}(2023)}]{Winckelmann2023}%
  \BibitemOpen
  \bibfield  {author} {\bibinfo {author} {\bibfnamefont {B.~G.}\ \bibnamefont
  {Winckelmann}}\ and\ \bibinfo {author} {\bibfnamefont {H.}~\bibnamefont
  {Bruus}},\ }\bibfield  {title} {\bibinfo {title} {Acoustic radiation force on
  a spherical thermoviscous particle in a thermoviscous fluid including
  scattering and microstreaming},\ }\href
  {https://doi.org/10.1103/PhysRevE.107.065103} {\bibfield  {journal} {\bibinfo
   {journal} {Phys. Rev. E}\ }\textbf {\bibinfo {volume} {107}},\ \bibinfo
  {pages} {065103} (\bibinfo {year} {2023})}\BibitemShut {NoStop}%
\bibitem [{\citenamefont {Muller}\ \emph {et~al.}(2012)\citenamefont {Muller},
  \citenamefont {Barnkob}, \citenamefont {Jensen},\ and\ \citenamefont
  {Bruus}}]{Muller2012}%
  \BibitemOpen
  \bibfield  {author} {\bibinfo {author} {\bibfnamefont {P.~B.}\ \bibnamefont
  {Muller}}, \bibinfo {author} {\bibfnamefont {R.}~\bibnamefont {Barnkob}},
  \bibinfo {author} {\bibfnamefont {M.~J.~H.}\ \bibnamefont {Jensen}},\ and\
  \bibinfo {author} {\bibfnamefont {H.}~\bibnamefont {Bruus}},\ }\bibfield
  {title} {\bibinfo {title} {A numerical study of microparticle acoustophoresis
  driven by acoustic radiation forces and streaming-induced drag forces},\
  }\href {https://doi.org/10.1039/C2LC40612H} {\bibfield  {journal} {\bibinfo
  {journal} {Lab Chip}\ }\textbf {\bibinfo {volume} {12}},\ \bibinfo {pages}
  {4617} (\bibinfo {year} {2012})}\BibitemShut {NoStop}%
\bibitem [{\citenamefont {Barnkob}\ \emph
  {et~al.}(2012{\natexlab{a}})\citenamefont {Barnkob}, \citenamefont
  {Augustsson}, \citenamefont {Laurell},\ and\ \citenamefont
  {Bruus}}]{Barnkob2012a}%
  \BibitemOpen
  \bibfield  {author} {\bibinfo {author} {\bibfnamefont {R.}~\bibnamefont
  {Barnkob}}, \bibinfo {author} {\bibfnamefont {P.}~\bibnamefont {Augustsson}},
  \bibinfo {author} {\bibfnamefont {T.}~\bibnamefont {Laurell}},\ and\ \bibinfo
  {author} {\bibfnamefont {H.}~\bibnamefont {Bruus}},\ }\bibfield  {title}
  {\bibinfo {title} {Acoustic radiation- and streaming-induced microparticle
  velocities determined by microparticle image velocimetry in an ultrasound
  symmetry plane},\ }\href {https://doi.org/10.1103/PhysRevE.86.056307}
  {\bibfield  {journal} {\bibinfo  {journal} {Phys. Rev. E}\ }\textbf {\bibinfo
  {volume} {86}},\ \bibinfo {pages} {056307} (\bibinfo {year}
  {2012}{\natexlab{a}})}\BibitemShut {NoStop}%
\bibitem [{Ond()}]{OndaCorpSolids}%
  \BibitemOpen
  \href@noop {} {\emph {\bibinfo {title} {Tables of Acoustic Properties of
  Materials: Solids}}},\ \bibinfo {organization} {Onda Corporation},\ \bibinfo
  {note} {\url{https://www.ondacorp.com/wp-content/uploads/2020/09/Solids.pdf},
  accessed 28 October 2025}\BibitemShut {NoStop}%
\bibitem [{\citenamefont {Muller}\ and\ \citenamefont
  {Bruus}(2014)}]{Muller2014}%
  \BibitemOpen
  \bibfield  {author} {\bibinfo {author} {\bibfnamefont {P.~B.}\ \bibnamefont
  {Muller}}\ and\ \bibinfo {author} {\bibfnamefont {H.}~\bibnamefont {Bruus}},\
  }\bibfield  {title} {\bibinfo {title} {Numerical study of thermoviscous
  effects in ultrasound-induced acoustic streaming in microchannels},\ }\href
  {https://doi.org/10.1103/PhysRevE.90.043016} {\bibfield  {journal} {\bibinfo
  {journal} {Phys. Rev. E}\ }\textbf {\bibinfo {volume} {90}},\ \bibinfo
  {pages} {043016} (\bibinfo {year} {2014})}\BibitemShut {NoStop}%
\bibitem [{\citenamefont {Ling}\ \emph {et~al.}(2020)\citenamefont {Ling},
  \citenamefont {Wei}, \citenamefont {Wang}, \citenamefont {Yang},
  \citenamefont {Qu},\ and\ \citenamefont {Fang}}]{Ling2020}%
  \BibitemOpen
  \bibfield  {author} {\bibinfo {author} {\bibfnamefont {B.}~\bibnamefont
  {Ling}}, \bibinfo {author} {\bibfnamefont {K.}~\bibnamefont {Wei}}, \bibinfo
  {author} {\bibfnamefont {Z.}~\bibnamefont {Wang}}, \bibinfo {author}
  {\bibfnamefont {X.}~\bibnamefont {Yang}}, \bibinfo {author} {\bibfnamefont
  {Z.}~\bibnamefont {Qu}},\ and\ \bibinfo {author} {\bibfnamefont
  {D.}~\bibnamefont {Fang}},\ }\bibfield  {title} {\bibinfo {title}
  {Experimentally program large magnitude of {Poisson}'s ratio in additively
  manufactured mechanical metamaterials},\ }\href
  {https://doi.org/10.1016/j.ijmecsci.2020.105466} {\bibfield  {journal}
  {\bibinfo  {journal} {Int. J. Mech. Sci.}\ }\textbf {\bibinfo {volume}
  {173}},\ \bibinfo {pages} {105466} (\bibinfo {year} {2020})}\BibitemShut
  {NoStop}%
\bibitem [{Com(2024)}]{Comsol62}%
  \BibitemOpen
  \href@noop {} {}\bibinfo {organization} {{COMSOL Multiphysics 6.2}} (\bibinfo
  {year} {2024}),\ \bibinfo {note} {\url{http://www.comsol.com}}\BibitemShut
  {NoStop}%
\bibitem [{\citenamefont {Ovchinnikov}\ \emph {et~al.}(2014)\citenamefont
  {Ovchinnikov}, \citenamefont {Zhou},\ and\ \citenamefont
  {Yalamanchili}}]{Ovchinnikov2014}%
  \BibitemOpen
  \bibfield  {author} {\bibinfo {author} {\bibfnamefont {M.}~\bibnamefont
  {Ovchinnikov}}, \bibinfo {author} {\bibfnamefont {J.}~\bibnamefont {Zhou}},\
  and\ \bibinfo {author} {\bibfnamefont {S.}~\bibnamefont {Yalamanchili}},\
  }\bibfield  {title} {\bibinfo {title} {Acoustic streaming of a sharp edge},\
  }\href {https://doi.org/10.1121/1.4881919} {\bibfield  {journal} {\bibinfo
  {journal} {J. Acoust. Soc. Am.}\ }\textbf {\bibinfo {volume} {136}},\
  \bibinfo {pages} {22} (\bibinfo {year} {2014})}\BibitemShut {NoStop}%
\bibitem [{\citenamefont {Barnkob}\ \emph
  {et~al.}(2012{\natexlab{b}})\citenamefont {Barnkob}, \citenamefont
  {Iranmanesh}, \citenamefont {Wiklund},\ and\ \citenamefont
  {Bruus}}]{Barnkob2012}%
  \BibitemOpen
  \bibfield  {author} {\bibinfo {author} {\bibfnamefont {R.}~\bibnamefont
  {Barnkob}}, \bibinfo {author} {\bibfnamefont {I.}~\bibnamefont {Iranmanesh}},
  \bibinfo {author} {\bibfnamefont {M.}~\bibnamefont {Wiklund}},\ and\ \bibinfo
  {author} {\bibfnamefont {H.}~\bibnamefont {Bruus}},\ }\bibfield  {title}
  {\bibinfo {title} {Measuring acoustic energy density in microchannel
  acoustophoresis using a simple and rapid light-intensity method},\ }\href
  {https://doi.org/10.1039/C2LC40120G} {\bibfield  {journal} {\bibinfo
  {journal} {Lab Chip}\ }\textbf {\bibinfo {volume} {12}},\ \bibinfo {pages}
  {2337} (\bibinfo {year} {2012}{\natexlab{b}})}\BibitemShut {NoStop}%
\bibitem [{\citenamefont {Hahn}\ and\ \citenamefont {Dual}(2015)}]{Hahn2015}%
  \BibitemOpen
  \bibfield  {author} {\bibinfo {author} {\bibfnamefont {P.}~\bibnamefont
  {Hahn}}\ and\ \bibinfo {author} {\bibfnamefont {J.}~\bibnamefont {Dual}},\
  }\bibfield  {title} {\bibinfo {title} {A numerically efficient damping model
  for acoustic resonances in microfluidic cavities},\ }\href
  {https://doi.org/10.1063/1.4922986} {\bibfield  {journal} {\bibinfo
  {journal} {Physics of Fluids}\ }\textbf {\bibinfo {volume} {27}},\ \bibinfo
  {pages} {062005} (\bibinfo {year} {2015})}\BibitemShut {NoStop}%
\bibitem [{\citenamefont {Karlsen}\ \emph {et~al.}(2018)\citenamefont
  {Karlsen}, \citenamefont {Qiu}, \citenamefont {Augustsson},\ and\
  \citenamefont {Bruus}}]{Karlsen2018}%
  \BibitemOpen
  \bibfield  {author} {\bibinfo {author} {\bibfnamefont {J.~T.}\ \bibnamefont
  {Karlsen}}, \bibinfo {author} {\bibfnamefont {W.}~\bibnamefont {Qiu}},
  \bibinfo {author} {\bibfnamefont {P.}~\bibnamefont {Augustsson}},\ and\
  \bibinfo {author} {\bibfnamefont {H.}~\bibnamefont {Bruus}},\ }\bibfield
  {title} {\bibinfo {title} {Acoustic streaming and its suppression in
  inhomogeneous fluids},\ }\href
  {https://doi.org/10.1103/PhysRevLett.120.054501} {\bibfield  {journal}
  {\bibinfo  {journal} {Phys. Rev. Lett.}\ }\textbf {\bibinfo {volume} {120}},\
  \bibinfo {pages} {054501} (\bibinfo {year} {2018})}\BibitemShut {NoStop}%
\bibitem [{\citenamefont {Winckelmann}\ and\ \citenamefont
  {Bruus}(2021)}]{Winckelmann2021}%
  \BibitemOpen
  \bibfield  {author} {\bibinfo {author} {\bibfnamefont {B.~G.}\ \bibnamefont
  {Winckelmann}}\ and\ \bibinfo {author} {\bibfnamefont {H.}~\bibnamefont
  {Bruus}},\ }\bibfield  {title} {\bibinfo {title} {Theory and simulation of
  electroosmotic suppression of acoustic streaming},\ }\href
  {https://doi.org/10.1121/10.0005051} {\bibfield  {journal} {\bibinfo
  {journal} {J. Acoust. Soc. Am.}\ }\textbf {\bibinfo {volume} {149}},\
  \bibinfo {pages} {3917} (\bibinfo {year} {2021})}\BibitemShut {NoStop}%
\bibitem [{\citenamefont {Tajima}\ \emph {et~al.}(2004)\citenamefont {Tajima},
  \citenamefont {Zhou}, \citenamefont {Nakajima},\ and\ \citenamefont
  {Sato}}]{Tajima2004}%
  \BibitemOpen
  \bibfield  {author} {\bibinfo {author} {\bibfnamefont {K.}~\bibnamefont
  {Tajima}}, \bibinfo {author} {\bibfnamefont {J.}~\bibnamefont {Zhou}},
  \bibinfo {author} {\bibfnamefont {K.}~\bibnamefont {Nakajima}},\ and\
  \bibinfo {author} {\bibfnamefont {K.}~\bibnamefont {Sato}},\ }\bibfield
  {title} {\bibinfo {title} {Ultralow loss and long length photonic crystal
  fiber},\ }\href {https://opg.optica.org/jlt/abstract.cfm?uri=jlt-22-1-7}
  {\bibfield  {journal} {\bibinfo  {journal} {Journal of Lightwave Technology}\
  }\textbf {\bibinfo {volume} {22}},\ \bibinfo {pages} {7} (\bibinfo {year}
  {2004})}\BibitemShut {NoStop}%
\bibitem [{\citenamefont {Bauer}\ \emph {et~al.}(2023)\citenamefont {Bauer},
  \citenamefont {Crook},\ and\ \citenamefont {Baldacchini}}]{Bauer2023}%
  \BibitemOpen
  \bibfield  {author} {\bibinfo {author} {\bibfnamefont {J.}~\bibnamefont
  {Bauer}}, \bibinfo {author} {\bibfnamefont {C.}~\bibnamefont {Crook}},\ and\
  \bibinfo {author} {\bibfnamefont {T.}~\bibnamefont {Baldacchini}},\
  }\bibfield  {title} {\bibinfo {title} {A sinterless, low-temperature route to
  {3D} print nanoscale optical-grade glass},\ }\href
  {https://doi.org/10.1126/science.abq3037} {\bibfield  {journal} {\bibinfo
  {journal} {Science}\ }\textbf {\bibinfo {volume} {380}},\ \bibinfo {pages}
  {960} (\bibinfo {year} {2023})}\BibitemShut {NoStop}%
\end{thebibliography}

%

\end{document}